\documentclass[sigconf]{acmart}
\usepackage{balance}
\usepackage{multirow}
\usepackage{multicol}
\usepackage{float}
\usepackage{subcaption}
\usepackage{graphicx}
\usepackage{enumitem}
\usepackage{color}
\usepackage{hyperref}
\hypersetup{
    colorlinks=true,      
    linkcolor=blue,
    filecolor=blue,      
    urlcolor=blue,
    citecolor=cyan,     
}

\newcommand*{\secref}[1]{\textbf{section \ref{#1}}}

\AtBeginDocument{%
  }

\setcopyright{acmcopyright}
\copyrightyear{2023}
\acmYear{2023}
\setcopyright{acmlicensed}

\acmConference[McGE '23] {Proceedings of the 1st International Workshop on Multimedia Content Generation and Evaluation: New Methods and Practice}{October 29, 2023}{Ottawa, ON, Canada.}
\acmBooktitle{Proceedings of the 1st International Workshop on Multimedia Content Generation and Evaluation: New Methods and Practice (McGE '23), October 29, 2023, Ottawa, ON, Canada}
\acmPrice{15.00}
\acmISBN{979-8-4007-0278-5/23/10}
\acmDOI{10.1145/3607541.3616821}

\begin{document}

\title{EMID: An Emotional Aligned Dataset in Audio-Visual Modality}

\author{Jialing Zou}
\authornote{Both authors contributed equally to this research.}
\email{51255901103@stu.ecnu.edu.cn}
\orcid{0009-0002-7583-0503}
\author{Jiahao Mei}
\authornotemark[1]
\email{10215102440@stu.ecnu.edu.cn}
\orcid{0009-0003-7544-4546}
\affiliation{%
  \institution{East China Normal University}
  \city{Shanghai}
  \country{China}
}
\author{Guangze Ye}
\affiliation{
  \institution{East China Normal University}
  \city{Shanghai}
  \country{China}
}
\email{5226591017@stu.ecnu.edu.cn}
\orcid{0009-0000-6831-5277}

\author{Tianyu Huai}
\affiliation{
  \institution{East China Normal University}
  \city{Shanghai}
  \country{China}
}
\email{tyhuai@stu.ecnu.edu.cn}
\orcid{0009-0000-3573-9962}

\author{Qiwei Shen}
\affiliation{
  \institution{East China Normal University}
  \city{Shanghai}
  \country{China}
}
\email{10215101495@stu.ecnu.edu.cn}
\orcid{0009-0000-9904-6233}

\author{Daoguo Dong}
\authornote{Corresponding author.}
\affiliation{
  \institution{East China Normal University}
  \city{Shanghai}
  \country{China}
}
\email{dgdong@cs.ecnu.edu.cn}
\orcid{0009-0009-7712-4734}

\renewcommand{\shortauthors}{Jialing Zou et al.}

\begin{abstract}
  In this paper, we propose \textbf{E}motionally paired \textbf{M}usic and \textbf{I}mage \textbf{D}ataset (\textbf{EMID}), a novel dataset designed for the emotional matching of music and images, to facilitate auditory-visual cross-modal tasks such as generation and retrieval. Unlike existing approaches that primarily focus on semantic correlations or roughly divided emotional relations, EMID emphasizes the significance of emotional consistency between music and images using an advanced 13-dimension emotional model. By incorporating emotional alignment into the dataset, it aims to establish pairs that closely align with human perceptual understanding, thereby raising the performance of auditory-visual cross-modal tasks. We also design a supplemental module named EMI-Adapter to optimize existing cross-modal alignment methods. To validate the effectiveness of the EMID, we conduct a psychological experiment, which has demonstrated that considering the emotional relationship between the two modalities effectively improves the accuracy of matching in abstract perspective. This research lays the foundation for future cross-modal research in domains such as psychotherapy and contributes to advancing the understanding and utilization of emotions in cross-modal alignment. The EMID dataset is available at \href{https://github.com/ecnu-aigc/EMID/tree/main}{\textcolor[RGB]{65,105,225}{https://github.com/ecnu-aigc/EMID.}}
\end{abstract}




\maketitle
\vspace{5pt}
\section{Introduction}
\label{sec:intro}
Guided Imagery and Music (GIM) is defined as a comprehensive therapy integrating professional knowledge of psychology, bio-medicine, and aesthetics \cite{bonny1989sound,bonny2010music}. Through associations and metaphors between music and images, patients can gain deeper insight into and explore their inner world, which helps to foster emotional expression and improve therapeutic results \cite{smyrnioti2023guided,agres2021music}. However, the current GIM scheme exhibits certain limitations, including imprecise targeting that lacks of pathological specificity, prolonged and labor-intensive procedures that result in high annotation costs, and so on.

Recently, the emergence of Artificial Intelligence Generated Content (AIGC) has instigated disruptive transformations in various cross-modal interaction scenarios. AIGC refers to a rising technology that utilizes advanced algorithms and deep learning methods to match or create a wide range of textual, visual, auditory, and multi-media content tailored to human demands \cite{cao2023comprehensive}. From an intuitive standpoint, AIGC offers a viable avenue to mitigate the costs associated with music and image collection and annotation in GIM. Furthermore, it has the potential to provide patients with a seamless and captivating audio-visual experience, ultimately enhancing the effectiveness of the treatment. Nevertheless, to the best of our knowledge, little existing research or implementation has been found that applies AIGC technology specifically to the field of psychotherapy.

\begin{figure}[t]
  \centering
  \vspace{10px}
  \includegraphics[width=\linewidth]{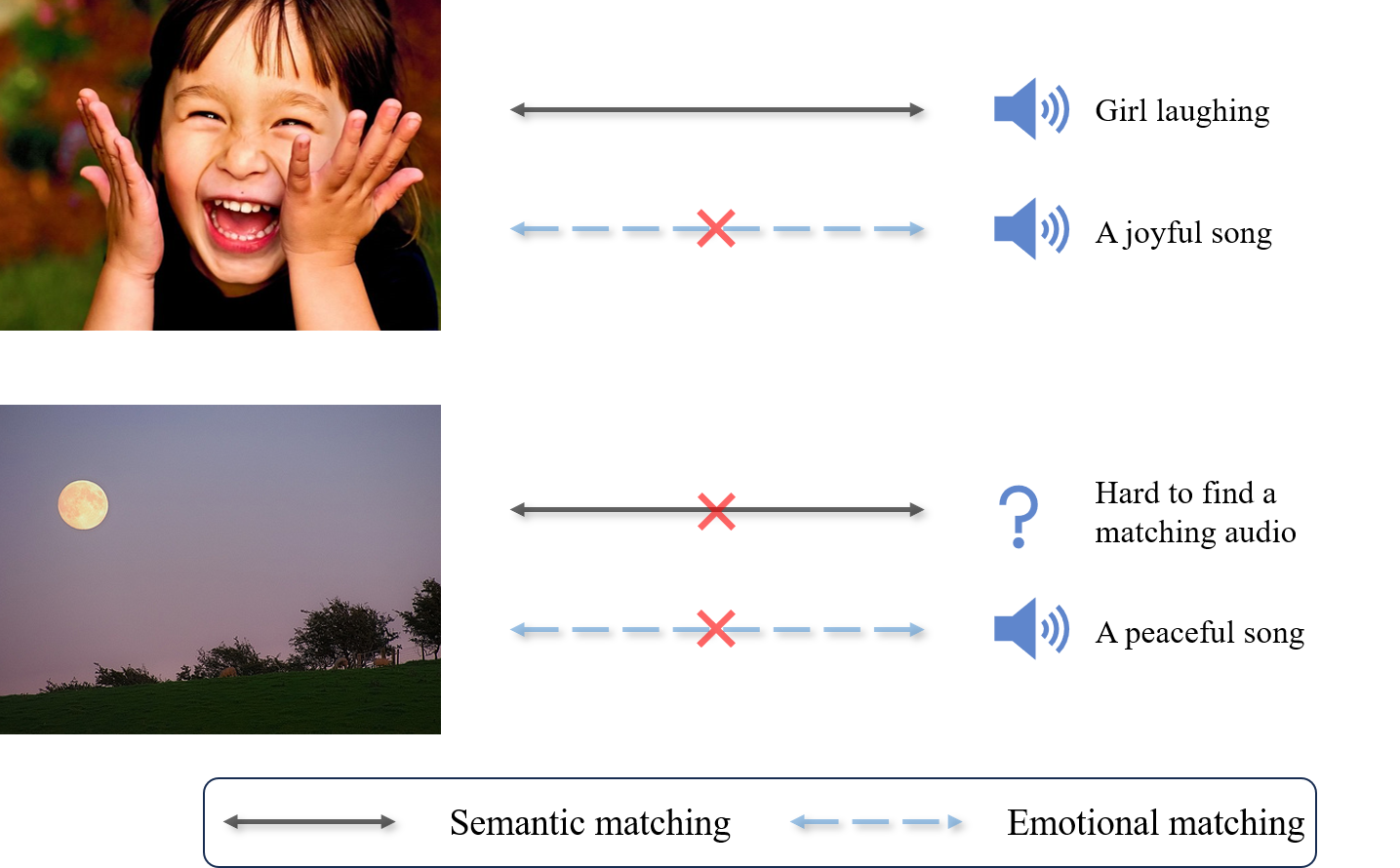}
    \caption{Defect of current cross-modal alignment approaches, ignoring the emotional relations of music and images. }
  \Description{Defect of current current cross-modal alignment approaches}
    \vspace{-10px}
  \label{fig:defect}
\end{figure}

Actually, the aforementioned phenomenon can be attributed to the current cross-modal alignment approaches relied on by AIGC, which primarily focus on semantic relation, ignoring emotional and artistic level matching, as shown in Figure~\ref{fig:defect}. An image of a laughing girl can easily correspond to an audio of giggling, but a photo of landscape without any sound clue may be challenging to access a suitable match at the semantic level. Meanwhile, it is challenging to find appropriate music that matches accordant artistic conception of both them without any explicit labels or guidance. Therefore, this article aims to propose a music-image dataset that incorporates emotional information, in order to fulfill the following objectives: (1) update the original cross-modal alignment pattern to project emotional features into the shared latent space. This adjustment is helpful to address challenges related to data collection and annotation reliance as well, (2) provide data support for training existing cross-modal generation and retrieval models. This dataset goes beyond simplistic and superficial semantic binding and takes into account the emotional associations between visual and auditory modalities, enhancing the applicability of the proposed dataset in the field of psychotherapy. 

To be more specific, our work makes the following key contributions: 
\begin{itemize}[leftmargin=13pt,itemsep=6pt,topsep=9pt]
\item We introduce a high-quality \textbf{E}motionally paired \textbf{M}usic and \textbf{I}mage \textbf{D}ataset (\textbf{EMID}), comprising more than 30k data pairs accompanied by abundant emotional annotations. 
\item We pioneer the attempt to align two distinct modalities based on emotions. Our approach involves extracting semantics from source data to form raw image-music pairs, subsequently determining the optimal matches by projecting each candidate onto the emotional coordinate space, thereby modifying embedded representations from original cross-modal joint embedding space.
\item A meticulously designed professional psychological experiment is conducted to assess the dataset's quality and affirm the pivotal role of emotions in modal alignment. 
\end{itemize}

Overall, the EMID will serve as a solid foundation for AIGC's application in psychotherapy-related work. Its availability paves the way for novel advancements in music and image conditional generation and retrieval, taking a closer step to human cognition and unlocking new possibilities and opportunities.

 \section{RELATED WORK}
\label{sec:related}

\subsection{Cross-modal Dataset}
The latest advancements in cross-modal dataset development labor primarily focus on obtaining images, audio, or videos based on textual inputs (text-to-image \cite{deng2009, lin2014microsoft,radford2021learning,schuhmann2021laion,schuhmann2022laion}; text-to audio/videos \cite{gemmeke2017audio,kim-etal-2019-audiocaps,9645159}) consisting of labels, descriptive information, evaluation content, etc., which are available throughout the internet and can be collected by automated tools with ease. The majority of text sections from existing cross-modal datasets are composed of limited tags or monotonous sentences \cite{nagrani2022learning}, hindering the establishment of strong sensory impacts through their collocation with images or audio. To overcome this limitation, efforts are being made to collect more impressive audio-visual combinations. Initially, videos and audio are often paired naturally, such as the huge dataset \text{AudioSet} \cite{gemmeke2017audio}, which is characterized as an object-oriented video dataset organized by audio events, using a meticulously constructed hierarchical ontology of 632 distinct audio classes, including at least 100 instances of 485 audio event classes. However, plain matching approach may lead to issues such as misalignment between video frames and audio, high data noise and classification errors \cite{9645159}.

Furthermore, in scenarios where audio is substituted with structured and creative music, the original semantic alignment scheme \cite{tang2023any} becomes obsolete. Nonetheless, various solutions have been proposed to address this issue. For example, MusicCaps \cite{agostinelli2023musiclm} manually annotated music segments of 5,521 pieces of \text{AudioSet} \cite{gemmeke2017audio}, each with an English text description written by ten professional musicians; Huang et al. \cite{huang2023noise2music} proposes a dataset called \text{MuLaMCap}, which leverages a pre-trained joint music-text model \text{MULAN} \cite{huang2022mulan} to annotate music segments in \text{AudioSet} \cite{gemmeke2017audio} based on music descriptions generated by LLM and 200k music descriptive sentences and labels from \text{MusicCaps} \cite{agostinelli2023musiclm}, retaining over 400k music-text pairs. The approach of emotional alignment, however, better conforms to human cognition \cite{li2019query,thao2023emomv}. Nevertheless, there have been rare research on the emotional alignment between images and music, with one of the solely known music-image paired dataset based on artistic conception, the \text{BOP} dataset \cite{fan2022conchshell}, which artificially annotates 3k+ images and 1.5k+ piano pieces based on creation background and emotions conveyed. Nonetheless, the manual annotation scheme obstructs data expansion, and the richness of source data is expected to be improved. Therefore, there is an urgent demand for a music-image matching dataset.

\begin{figure*}[t]
    \centering
    \includegraphics[width=0.98\linewidth]{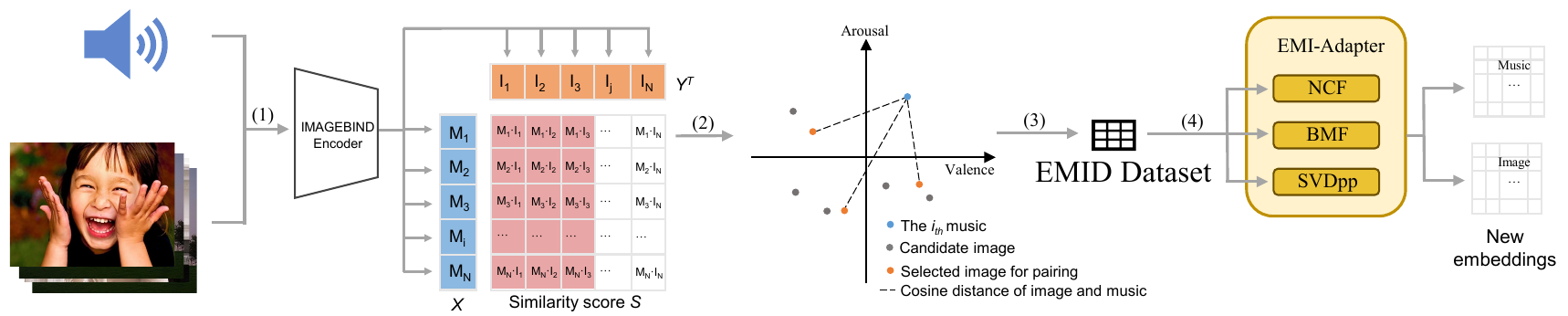}%
    \caption{The construction process of EMID includes the following steps. (1) Use ImageBind to get the semantic embedding for images and music as matrix $X$ and $Y$. Then we get the similarity score matrix through $S = Softmax(X\cdot{Y^{T}})$, (2) select $K$ images with the highest scores (in red color) from $i_{th}$ row of $S$ as the candidate image-set for the $i_{th}$ music, then calculate the cosine distance of each music clip and its image-candidates in emotional valence-arousal space, (3) generate EMID through weighted summation of semantic scores and emotional scores, remaining $m$ best-matching pairs, (4) design a lightweight component, EMI-Adapter, which can merge emotional features into the joint embedding space after training based on the EMID.}
    \label{fig:pipeline}
\end{figure*}

\subsection{Cross-modal Task}
Cross-modal tasks involving generation, retrieval, and editing are predominantly addressed using state-of-the-art frameworks such as GAN \cite{goodfellow2020generative}, VQ-VAE \cite{van2017neural,razavi2019generating}, and Diffusion models \cite{ho2020denoising,rombach2022high}. The recent rapid development of AIGC has garnered substantial attention from both academia and industry. Research efforts \cite{lee2022sound,li2022learning,lee2023soundini} have been directed towards visual editing guided by sound and vice versa, taking advantage of conditional guidance to facilitate tasks like style transfer, local fine-tuning, inpainting and super-resolution while preserving the integrity of the original content. 

Regarding generative work, there is still ample room for advancement. Early generative works focus on restoring sound of silent instrument-playing videos based on the performer's body movements or gestures \cite{su2020audeo,gan2020foley}. Subsequent jobs attempt to generate corresponding audio based on images or videos of animals, water flow, knocking actions and more \cite{zhou2018visual,sheffer2023hear,ruan2023mm}. However, this type of audio often amounts to a semantic repetition of the image, lacking imaginative and artistic audio-visual elements. To overcome this constraint, recent works have explored novel manners to introduce more imaginative and artistic ideas into the generation process. Fan et al. \cite{fan2022conchshell} designs a multi-modal architecture named ConchShell based on GAN networks \cite{goodfellow2020generative}, which can generate piano music clips of around 8 seconds conditioned on input landscape images. Surís et al. \cite{suris2022s} propose a new method for recommending music tracks that align with a given video (or vice versa), without extra reliance on manual annotation, establishing a connection between the audio-visual pairs at a rhythmic and artistic level.

Training multi-modal data representations that can be aligned in latent spaces has been proved to offer robust support for conditional generation tasks \cite{liu2022cross,girdhar2023imagebind,tang2023any}. However, there is still a giant research gap in the domain of direct emotional alignment specifically focused on music and images. Therefore, the \textbf{EMID} proposed in this article aims to lay the foundation for this specific aspect of research and development.

\section{The Proposed Dataset: EMID}
\label{sec:method}

We construct the \textbf{EMID}, a high-quality collection of music-image pairs taking account of affective aspects. The EMID ensures a semantic closeness between music and images while evoking similar affective experiences and establishing a unified space of imagination. In the following sections, we will provide a comprehensive and detailed description of the emotional alignment process. 
\subsection{Emotional Alignment Process}
The automated construction process of the EMID involves three distinct phases: (1) acquisition of emotional features, (2) formation of candidate pairs, and (3) injection of emotional information. The overall pipeline is illustrated in Figure~\ref{fig:pipeline}. 
\subsubsection{Acquisition of Emotional Features}
Both images and music have the ability to convey intricate emotions, although their methods of evoking feelings and intensities differ. As a first step, we aim to extract comparable emotional information from individual samples of music and images. Each music clip conveys a distinctive emotion. For example, the combination of adagio tempo, deep tones, subdued harmonies, and gentle rhythms can evoke a poignant sense of sadness (e.g. the second movement of Dvorak's \emph{New World symphony}). On the other hand, fast rhythms, strong percussion, and lively melodies can convey feelings of happiness and joy (e.g. the first movement of Bach's \emph{Brandenburg Concerto}).

To acquire a diverse range of emotional data, we refer to the work conducted by Cowen et al. \cite{cowen2020music}, from whose visualization web page\footnote{\url{https://www.ocf.berkeley.edu/~acowen/music.html}} we crawl 1,841 music clips, each accompanied by rich emotional annotations. These music clips encompass various genres, including rock, folk, jazz, classical, wind, and heavy metal, covering a broad spectrum of emotional dimensions, such as amusing, annoying, anxious/tense, beautiful, calm/relaxing, dreamy, energizing/pump-up, erotic/desirous, indignant/defiant, joyful/cheerful , sad/depressing, scary/fearful and triumphant/heroic that contains 13 emotions in total, which are represented by letters $A$ to $M$ in Figure~\ref{fig:EMID-hotmap}, Figure~\ref{fig:pairAccuracyVs.emotion} and Table~\ref{tab:quantity}.

\begin{figure*}[t]
  \centering
  
  \subfloat[For each of the eight emotional categories, we select three semantically similar words from the NRC-VAD-Lexicon dictionary as emotional descriptions, which are marked with the same color in the figure. Note that the range of Valence-Arousal values is \begin{math}(0,1)\end{math}. ]{%
    \includegraphics[width=0.45\linewidth]{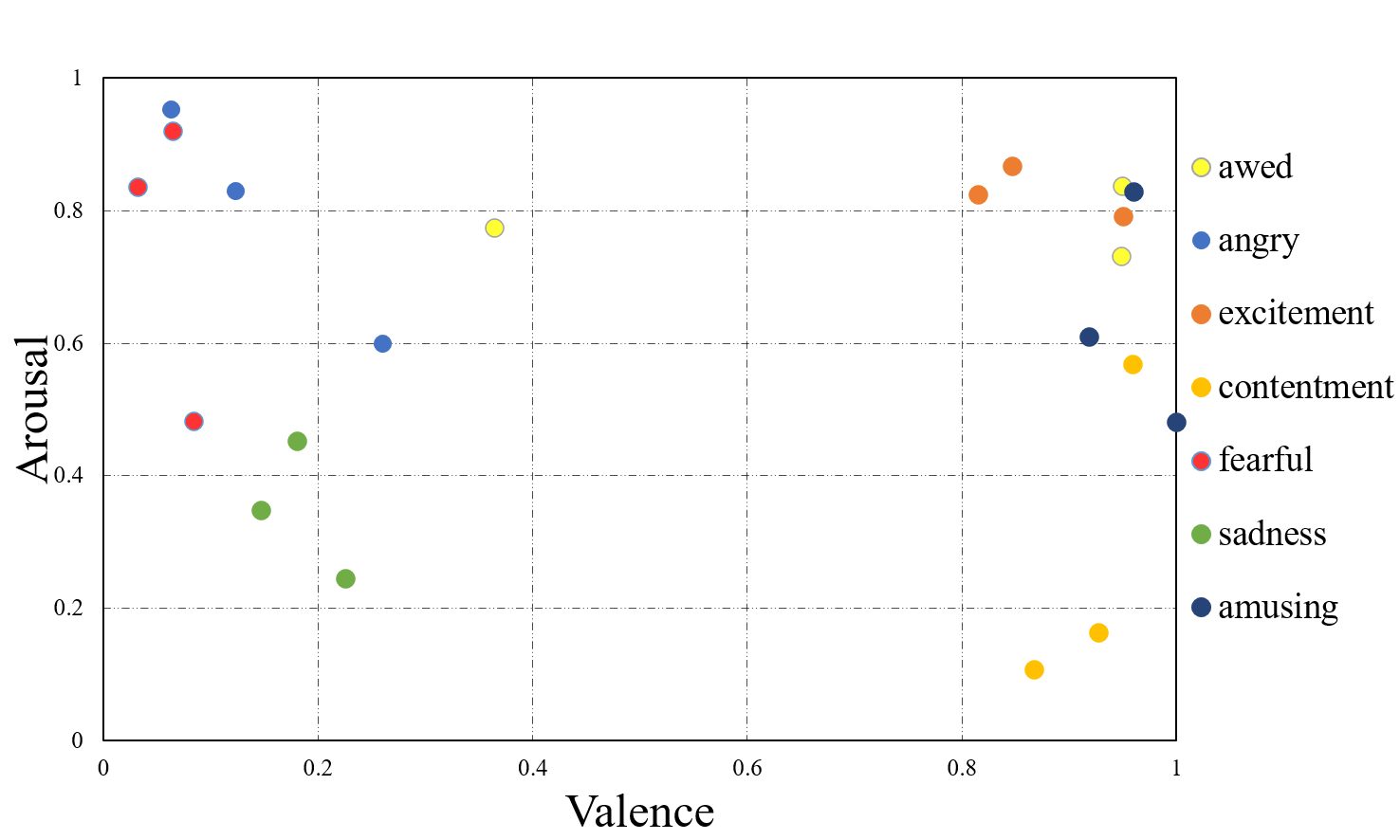}%
    \label{fig:va_left}%
    
  } 
  \hfill
  \subfloat[To obtain the approximate Valence-Arousal coordinates for every image category, we calculate the mean values of the emotional description coordinates in Figure~\ref{fig:va_left} for them and map these values to \begin{math}(-1,1)\end{math}. ]{%
    \includegraphics[width=0.45\linewidth]{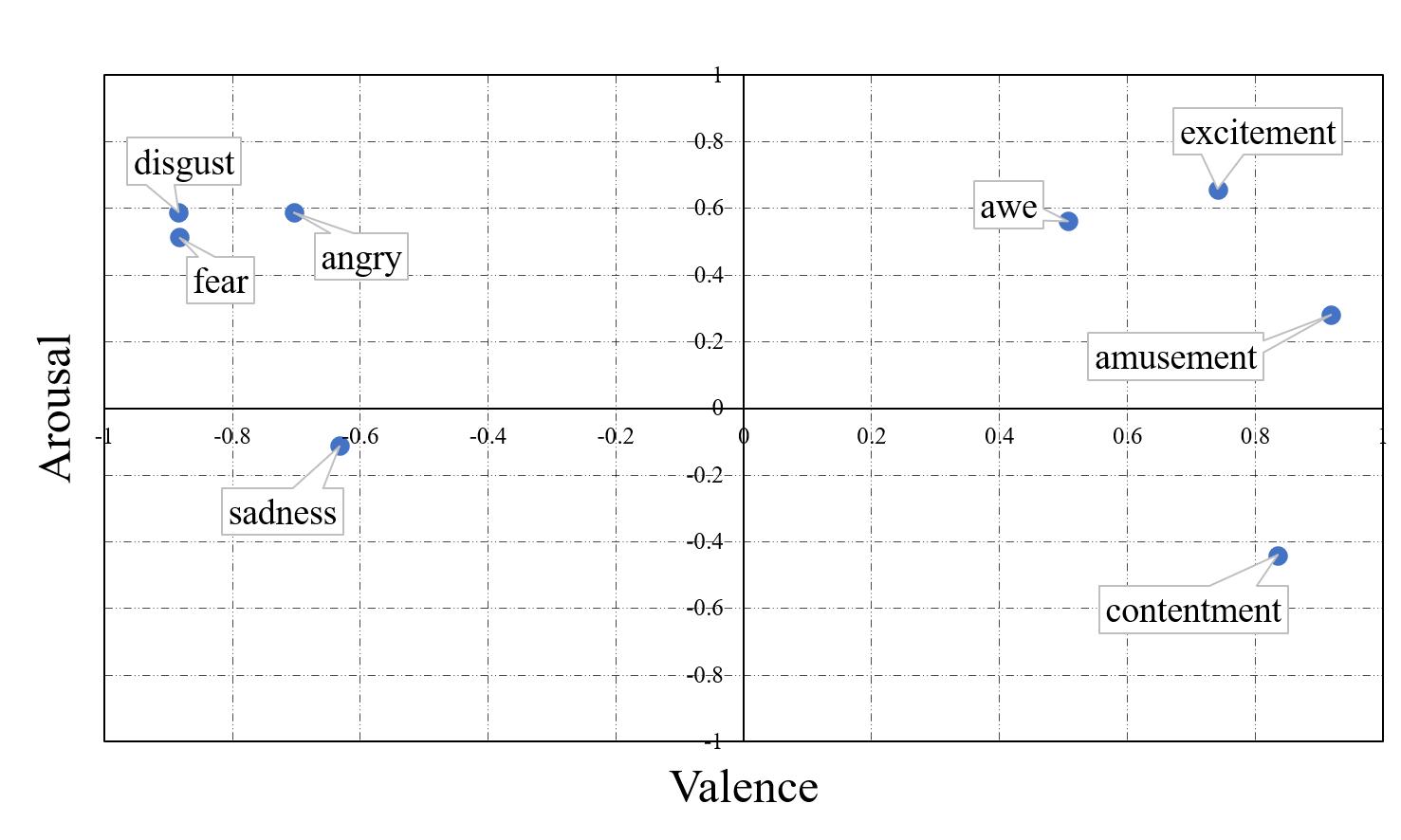}%
    \label{fig:va_right}%
  }
  
  \caption{Valence-Arousal coordinates of eight emotional categories from the NRC-VAD-Lexicon \cite{mohammad2018obtaining} dictionary and the final calculated values}
  \label{fig:va}
\end{figure*}

To establish a standardized representation of emotions, we adopt the Valence-Arousal space theory proposed by Hanjalic \cite{hanjalic2005affective}. In this theory, valence is used to measure the positive or negative aspects of music emotions, while arousal describes the level of emotional stimulation. For images, we utilize the 8-dimensional emotion classification image data proposed by You et al. \cite{you2016building}. This classification system comprises eight categories: happiness, awe, contentment, surprise, anger, disgust, fear, and sadness. To establish an emotional connection between music and images, we exert the NRC-VAD-Lexicon \cite{mohammad2018obtaining} dictionary to select three semantically similar words for each image classification, which serve as emotional descriptions for that particular category. Subsequently, we query the dictionary to obtain the \textit{<V, A>} coordinates corresponding to these eight classifications (shown in Figure~\ref{fig:va_left}) and calculate their averages as the approximate \textit{<V, A>} coordinates for the images within each category (shown in Figure~\ref{fig:va_right}).

It is worth noting that individual images are not specifically assigned with independent coordinates, as the presence of classification labels already constrain an estimated range of each sample within the Valence-Arousal space. By employing the classification coordinates to indicate the location of sample points, we dramatically reduce the computational expenses associated with measuring emotional distances in subsequent analysis, while also minimizing any unfavorable influence on the ultimate matching efficacy.

\subsubsection{Formation of Candidate Pairs}
Based on previous work, we initially pick several matched images for each music sample based on semantic similarities to create an initial dataset. Specifically, we employ the cross-modal retrieval model ImageBind \cite{girdhar2023imagebind} to calculate the semantic matrices for both music and image datasets, recording as $X \in R^{N_M*d}, Y \in R^{N_I*d}$ respectively, Where $N_M, N_I$ refer to the number of music and image samples, and $d$ represents the dimension of latent space that pertains to ImageBind. By performing matrix multiplication, we attain a similarity score between each music sample and image sample, resulting in a semantic similarity matrix \textit{S}. The specific calculation process is as follow:
\begin{equation}
S = Softmax(X\cdot{Y^{T}})
\end{equation}

Each element $S_{ij} = \frac{M_i\cdot I_j}{\|M_i\|\cdot \|I_j\|}$ reflects the semantic distance between the $i_{th}$ music segment and the $j_{th}$ image. Utilizing this matrix, we select $K$ (setting $K= 100$ in the EMID) images with the highest semantic relevance for each music sample, thereby setting up candidate music-image pairs. Note that this raw dataset solely relies on simple semantic associations, which may result in suboptimal matching performance. To further tackle this issue, we consider the integration of emotional information.

\subsubsection{Injection of Emotional Information}
\label{3.1.3}
 Finally, we adopt a similar method as above to calculate the \textit{<V, A>} coordinates of each music sample  and the cosine distance from images in its candidate set in Valence-Arousal space so as to acquire the emotional similarity matrix $E$ between music and alternative images. Then we add semantic and emotional similarity matrices by adjusting appropriate weight parameters $\alpha$ to generate the final score matrix $T$ for pairing as follows:
 \begin{equation}\label{H}
 T = \alpha \cdot S + (1 - \alpha) \cdot E
 \end{equation}
 This matrix determines the compatibility between music and image samples. To produce a set of $N_M * m$ music image pairs, we choose $m$ images for each music sample based on the score ranking from highest to lowest. 
 
\begin{figure*}[t]
  \centering
  \subfloat[Curve of training loss]{%
    \includegraphics[width=0.3\linewidth]{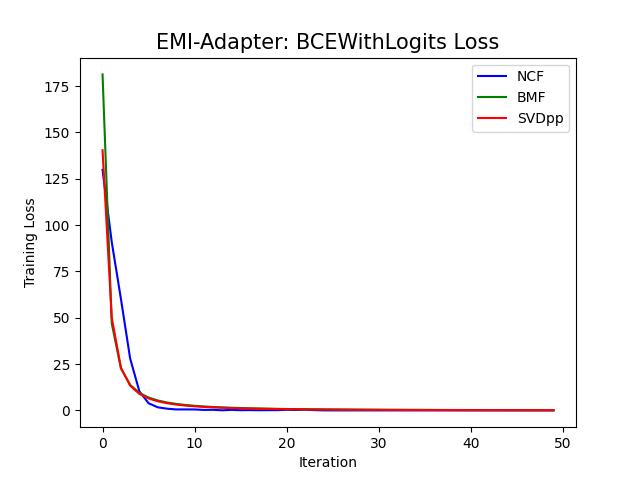}%
    \label{fig:EMI-loss}%
    
  } 
  \hfill
  \subfloat[Curve of HR scores]{%
    \includegraphics[width=0.3\linewidth]{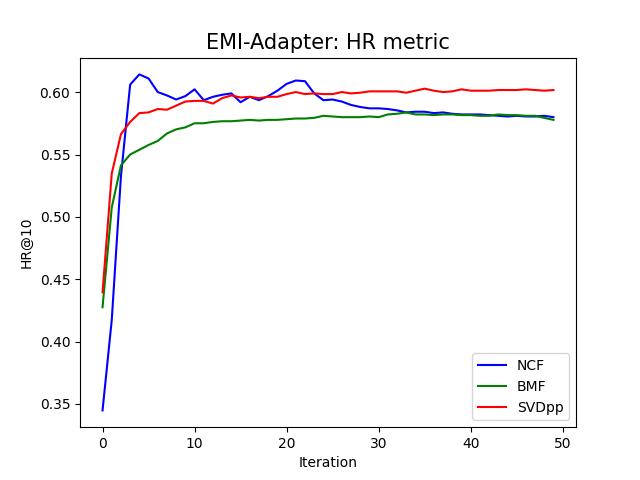}%
    \label{fig:EMI-HR}%
  }
    \hfill
  \subfloat[Curve of NDCG scores]{%
    \includegraphics[width=0.3\linewidth]{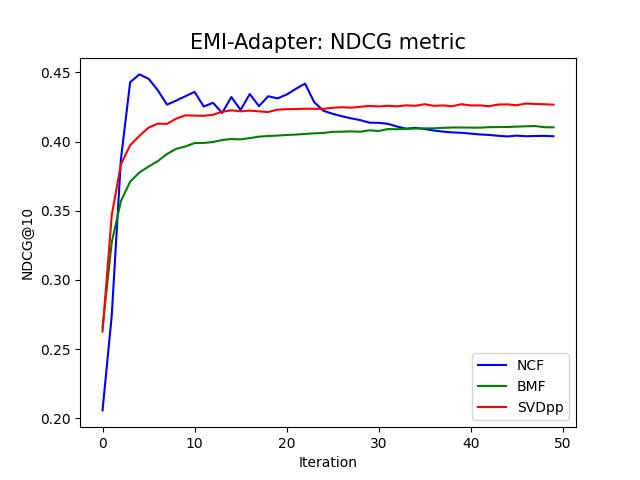}%
    \label{fig:EMI-NDCG}%
  }
  
  \caption{The above curves display the variation of training loss and evaluation results in 50 iterations for three types of networks, with $@10$ indicating that top-10 images will be recommended. At this point, the EMID version used for training is taken as $m = 3$ and $\alpha = 0.6$.}
  \label{fig:EMI-Adapter}
  \vspace{5pt}
\end{figure*}

 To ensure that the newly matched samples are as close as possible in the embedding space, a tiny emotional fusion plugin, \textit{EMI-Adapter}, is designed to regulate the encoding of ImageBind and incorporate emotional information into the shared latent space. We simultaneously train several lightweight network modules, comprising of NCF \cite{he2017neural}, BMF \cite{koren2009matrix}, and SVDpp \cite{koren2008factorization}, as adapters to promote cross-modal alignment based on existing music-image pairs. Results present in Table~\ref{tab:EMI-Adapter} and Figure~\ref{fig:EMI-Adapter} illustrate the performance of the different EMI-Adapter plugins.

Moreover, the EMI-Adapter has the capability to capture more comprehensive emotional information from the input modes leveraging the foundation of ImageBind, eliminating the reliance on additional emotional annotations provided by the source samples. This significant advancement will greatly facilitate the future expansion of the EMID, enabling it to encompass a broader range of emotional content.

 \begin{table}[t]
  \vspace{5pt}
  \caption{Performance of different EMI-Adapters. The best results are viewed in bold.}
  \label{tab:EMI-Adapter}
  \begin{tabular}{c|cc}
    \toprule
        & HR@10 & NDGC@10 \\
    \midrule
    NCF & 0.580 & 0.404\\
    BMF & 0.580 & 0.410\\
    SVD & \textbf{0.602} & \textbf{0.427}\\
    \bottomrule
  \end{tabular}
  \vspace{-2pt}
\end{table}

\begin{table*}[t]
\centering
\caption{The quantities of music dataset before/after expanding and the average of manually set thresholds on 13 emotions.  }
\label{tab:quantity}
\begin{tabular}{c|cccccccccccccc}
\toprule
Emotion & \textbf{A} & \textbf{B} & \textbf{C} & \textbf{D} & \textbf{E} & \textbf{F} & \textbf{G} & \textbf{H} & \textbf{I} & \textbf{J} & \textbf{K} & \textbf{L} & \textbf{M} & \textbf{Total} \\ \midrule
Before expanding& 45         & 80         & 54         & 131        & 306        & 174        & 367        & 86         & 36         & 124        & 129        & 105        & 199        & 1836           \\ 
Average threshold                           & 0.611      & 0.606      & 0.559      & 0.620      & 0.707      & 0.632      & 0.751      & 0.661      & 0.610      & 0.669      & 0.603      & 0.570      & 0.628      & -              \\ 

After expanding                        & 255        & 545        & 320        & 771      & 1531       & 889        & 1832       & 1036        & 323        & 1014        & 799        & 484        & 939        & 10738          \\  \bottomrule
\end{tabular}
\vspace{5pt}
\end{table*}

\subsection{Filtering and Expansion of Materials}
\subsubsection{Data Filtering}
To keep data balance, we exclude several music samples with a duration of less than 3.7 seconds from the originally crawled music pieces, leaving behind a final dataset of 1,836 music samples with associated tags, among which $98\%$ of the music clips held a duration of 4.8s to 5.2s, as the longest is 9.4s and the shortest is 3.7s. Regarding the image dataset, for the purpose of psychotherapy, we decide to exclude the \textit{disgusting} category, which contains elements of threat, violence, and blood. In addition, to avoid the negative impact of image quality on pairing, we manually define thresholds for the mean and variance of image brightness. Images falling below these thresholds are considered invalid or of low quality and then filtered out. Furthermore, we employ the GIT model \cite{wang2022git} to generate text descriptions for all images. Based on these captions, we remove images that are primarily characterized by textual information rather than visual content, as well as photos of bands or singers performing on stage as far as possible, for the sake of escaping from the impediment of strong semantics to emotional connection. After these filtering steps, the final retained image dataset consists of 19,901 pieces.

\subsubsection{Data Expansion}
To gather more available data, we expand the music samples by leveraging the API of the music listening and recognition platform\footnote{\url{https://audd.io/}}. This allows us to obtain 1,361 integral music tracks corresponding to fragments in our original dataset. Next, using the \textit{librosa} library\footnote{\url{https://github.com/librosa/librosa}}, we  compute the cross-similarity matrix between the music clips in the existing dataset and original music tracks by employing a sliding window approach with a length of 5 seconds and a step size of 1 second. Based on the idea of dynamic programming, we obtain the similarity score (ranged from $0$ to $1$) from the cross-similarity matrix. Subsequently, we determine the selection of new music clips by setting a manually defined threshold to the similarity score of each music, primarily aiming to differ from the original ones in terms of auditory content while reserving similar affectivity. These new clips are assigned the same labels as the corresponding source samples. The quantities of the music dataset before and after expansion are presented in Table~\ref{tab:quantity}, and the ultimately expanded music dataset involves 10,738 music clips.

\subsection{Dataset Statistics}
The ultimate dataset comprises 32,214 pairs of images and music (setting $m = 3$). The EMID is accompanied with rich annotations as the music samples are equipped with three types of labels: genre, emotion, and sentiment, among which the genre labels classify the music into 13 emotional dimensions, while the sentiment labels provide a detailed breakdown of the percentage distribution of each music sample across different emotional dimensions, and the emotion labels embody another 11 sensory dimension values that quantify the subjective experience evoked by the music. On the other hand, the image samples are marked with category labels and brief content descriptions. This information assists in further understanding the intricate connections between the two modalities. The location of pairs from the EMID in the cross-emotional space of music and images before and after considering emotional alignment is shown in the following Figure~\ref{fig:EMID-hotmap}. It can be observed that the distribution of data is more balanced on different emotional dimensions after fusing effective information.
\begin{figure*}[t]
  \centering
  \subfloat[]{%
    \includegraphics[width=0.47\linewidth]{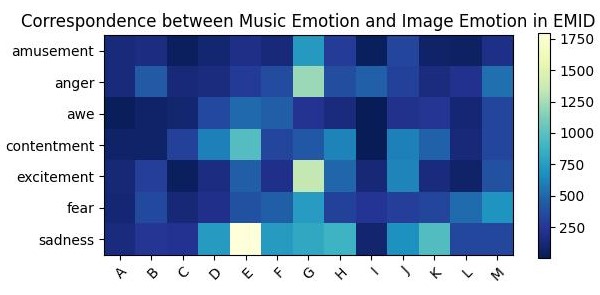}%
    \label{fig:EMI-before}%
    
  } 
  \hfill
  \subfloat[]{%
    \includegraphics[width=0.47\linewidth]{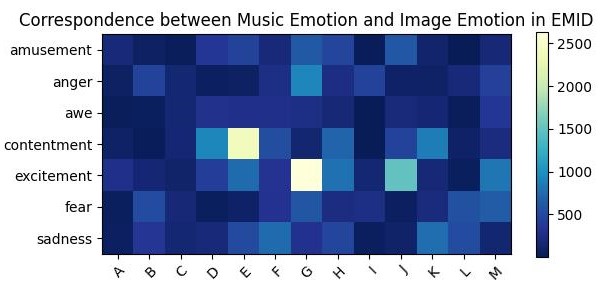}%
    \label{fig:EMI-after}%
  }
  
  \caption{Distribution of music-image pairs in the EMID without ($\alpha = 1$) and with ($\alpha = 0.6$) emotional alignment.}
  \label{fig:EMID-hotmap}
\end{figure*}

\section{ASSESSMENT Experiment}
\label{experiments.tex}

\begin{table}[t]
  \caption{Accuracy of Music-Image Matching Study. The best results are viewed in bold.}
  \label{tab:commands}
  \begin{tabular}{c|cccc}
    \toprule
    \hline
    & \textbf{Max} & \textbf{Min} & \textbf{Mean} & \textbf{Variance}\\
    \midrule
    &\multicolumn{4}{c}{\textit{Random Match}}\\
    \midrule
    Music-to-Image & 0.363 & 0.176 & 0.252 & 0.002\\
    Image-to-Music & 0.341 & 0.132 & 0.247 & 0.002\\
    Overall & 0.330 & 0.176 & 0.250 & 0.001\\
    \midrule
    \midrule
    &\multicolumn{4}{c}{\textit{Before Emotional Aligned}}\\
    \midrule
    Music-to-Image & 0.604 & \textbf{0.429} & 0.516 & 0.005\\
    Image-to-Music & \textbf{0.725} & \textbf{0.462} & \textbf{0.575} & 0.012\\
    Overall & 0.621 & 0.445 & 0.546 & 0.005\\
    \midrule
    \midrule
    &\multicolumn{4}{c}{\textit{After Emotional Aligned (Ours)}}\\
    \midrule
    Music-to-Image & \textbf{0.648} & 0.407 & \textbf{0.556} & 0.003\\
    Image-to-Music & 0.692 & 0.352 & 0.563 & 0.011\\
    Overall & \textbf{0.659} & \textbf{0.462} & \textbf{0.560} & 0.005\\
    \bottomrule
  \end{tabular}
\end{table}

\subsection{Experimental Protocol}
\textbf{Preview.} To assess the efficacy of the EMID, we conduct two types of psychological experiments: music-to-image validation and image-to-music validation. In order to explore whether intensifying emotional connections can effectively support cross-modal alignment research, subjects are instructed to determine one optimal match with the assigned example from the other modality according to their subjective intuition,

\subsubsection{Subjects}
The study involves 50 subjects from East China Normal University, comprising 29 males and 21 females. The average age of the subjects is approximately 21 years old. All subjects are right-handed, with normal naked or corrected vision. 
\subsubsection{Design}
Music-to-Image Verification. Each subject is supposed to complete 91 \textit{(7 × 13)} trials as 7 music clips from 13 emotional categories are selected respectively as stimulus materials. In each trial corresponding to one music clip, subjects are informed to choose the most suitable image from the four options based on their personal feeling of music expression after listening to the music clip. Only one of the options is the correct answer determined by the EMID (i.e. the best match estimated from the score matrix in \secref{3.1.3}), and the remaining three images are considered as interference items. All options belong to different emotional categories.

Image-to-Music Verification. Each subject is supposed to complete 91 \textit{(7 × 13)} trials as 7 sets of matched images from 13 emotional categories are selected respectively as stimulus materials. In each trial corresponding to one image set, subjects will be presented with three images used for describing the same music clip (i.e. $m$ images that ultimately pair with the music clip in \secref{3.1.3}, setting $m = 3$). After observing three images, subjects are informed to choose the most suitable music clip from the four options based on their personal feeling of image expression. The settings for correct answers and interference items are the same as above.

\begin{figure}[t]
  \centering
  \includegraphics[width=\linewidth]{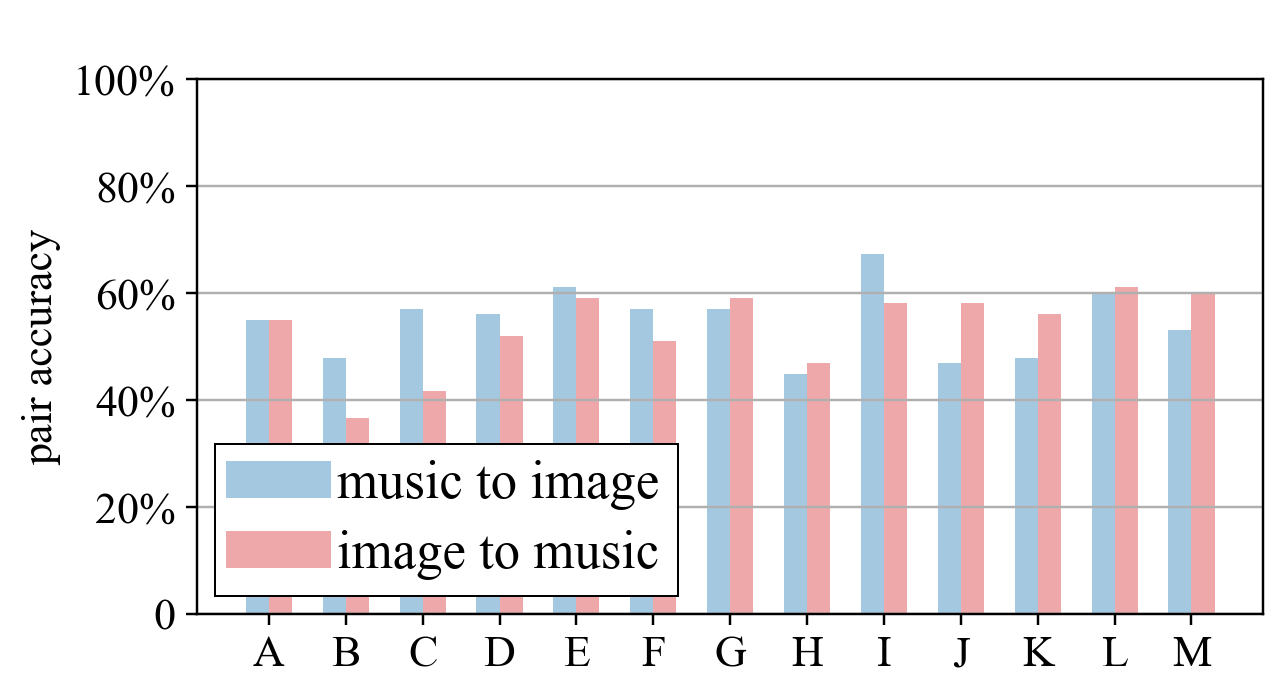}
  \caption{Pair accuracy of Image-to-Music Verification and Music-to-Image Verification on the 13 emotions, letter A to M stands for different emotions. }
  \Description{pairAccuracyVs.emotion}
  \label{fig:pairAccuracyVs.emotion}
\end{figure}

\textbf{Ablation Experiment}. To further probe the effect of emotions on alignment work, we also design an additional small-scale ablation experiment for comparison. To be more specific, we build a variant of the EMID by setting the $\alpha = 1$ in formula \ref{H}, thus we attain a dataset merely considering ImageBind semantic matching. By adopting the same partitioning scheme utilized in the previous experiment, we can compare the impact of image pairing before and after emotional alignment on final accuracy while maintaining consistent music coverage. A total of 15 subjects participated in this experiment.

Additionally, in order to verify that semantic alignment also plays a crucial role as the first step of our work, we simulate random matching scenarios for two types of validations (actually achieved by randomly generating user responses). The interference term settings in the ablation experiment remain consistent with the previous experiment.

\subsubsection{Evaluation}
Based on the response data of different subjects, we calculate the separate accuracy and overall accuracy of music-to-image and image-to-music experiments. Additionally, we conduct an analysis to investigate the distribution of accuracy across different music emotional categories. Note that accuracy refers to the ratio of the number of consistent situations between the subjects' answers and pairs in the EMID to the total number of validation questions.

\subsection{Performance Analysis}
The statistical information of the subject data collected is presented in Table \ref{tab:commands}. The results show that the music-image matching with emotional alignment outperforms other schemes in overall validation. To be more specific, simulated random matching conforms to our conjecture, with an overall mean accuracy of 0.25. Semantic matching based on ImageBind performs significantly better than random pairing, with a variance of 0.005, the maximum accuracy is 0.621, the minimum is 0.445, and the mean is 0.546. After emotional alignment, the matching accuracy improves further in both music-to-image validation and overall experiment, as the general assessment has a variance of 0.005, with a maximum accuracy of 0.659, a minimum of 0.462, and a mean of 0.560. However, in image-to-music verification, the effect after emotional alignment is slightly inferior to before alignment. This may be due to the strong semantics presented in the three images selected through ImageBind, which can definitely correspond to specific background sounds or instruments in the music. But in music-to-image validation which picks one suitable image based on a more abstract form of music, both semantic and emotional factors will be taken into account as images and music used in psychotherapy are supposed to evoke strong emotions rather than having explicit semantics.

Figure~\ref{fig:pairAccuracyVs.emotion} presents that the accuracy distribution of music-image matching in the EMID tends to average as a whole in different music emotion categories. In some categories, there are notable differences in accuracy between different validation directions, particularly in categories $C$ (anxious/tense) and $J$ (joyful/cheerful). In the case of the former category, the accuracy of music-to-image verification is significantly higher compared to the other direction. This difference may be attributed to the removal of the \textit{disgusting} classification in the source image dataset that is closest to category $C$ at the sensory level. On the other hand, the latter is exactly the opposite, possibly due to the large scale of \textit{amusement} classification in the source image dataset, which possesses similar feelings to category $J$. 

Furthermore, we notice that the validation accuracy of category $I$ (indignant/defiant) is relatively higher, while category $B$ (annoying) and category $H$ (erotic/desirous) tend to have lower accuracy. This pattern may be influenced by the incomplete equilibrium distribution of the source dataset. Additionally, the interference between similar emotions may also affect the accuracy of classification. To gain broader insights into these issues, our future research will further explore the reasons behind these accuracy differences and addressing these challenges.

\section{Conclusion}
\label{{conclusions.tex}}

In this paper, we proposed an emotional aligned music-image paired dataset named EMID, which considers the sentimental connections of different modalities in addition to semantic relations. Moreover, we enhanced the existing cross-modal alignment approaches through a simple module called EMI-adapter, which contributes to effectively capturing and integrating emotional information from the indicated samples and projecting it into the joint embedding space. In the end, the psychological experiment demonstrated that the matching pattern in the EMID is more in line with human cognition compared to traditional schemes, and also more suitable for cross-modal tasks in the field of psychotherapy. Currently, our dataset still suffers issues like insufficient scale, weak music structure, and limited data coverage. These challenges will be further discussed and addressed in our future research.


\bibliographystyle{ACM-Reference-Format}
\balance
\bibliography{sample-base}





\end{document}